\documentclass[aps,eqsecnum,groupedaddress,a4paper,twocolumn,floatfix,footinbib]{revtex4}
\usepackage[dvips]{graphicx}
\usepackage[T1]{fontenc}
\usepackage{mathptmx}
\usepackage{fancyhdr}
\newcommand{\XDOI}[1]{\href{http://dx.doi.org/#1}{doi:#1}}

\begin{document}

\title{Towards quantum well hot hole lasers}

\author{P. Kinsler}
\email{Dr.Paul.Kinsler@physics.org}
\author{W. Th. Wenckebach}
\affiliation{Department of Applied Physics,
Faculty of Applied Sciences,
T.U. Delft, \\
Lorentzweg 1,
2628 CJ Delft,
The Netherlands.}

\renewcommand{\baselinestretch}{1.00}

\lhead{\includegraphics[height=5mm,angle=0]{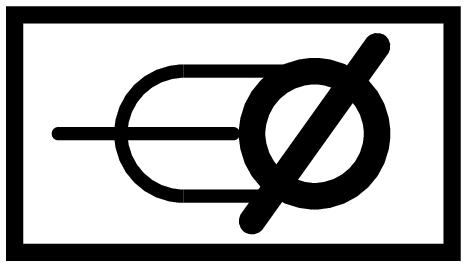}~~HHQWELL}
\chead{Towards QW hot hole lasers}
\rhead{
\href{mailto:Dr.Paul.Kinsler@physics.org}{Dr.Paul.Kinsler@physics.org}\\
\href{http://www.kinsler.org/physics/}{http://www.kinsler.org/physics/}
}



\begin{abstract}

 It should be possible to improve hot-hole laser performance by moving from
bulk materials to a quantum well structure. The extra design parameters enable
us to alter the band structure by changing the crystal orientation of the
growth direction; to use the well width to shift the subband offsets, enabling
the effect of the LO phonon scattering cut-off to be controlled; and to use
modulation doping to ensure a high hole concentration to increase the gain
without the dopants being present in the gain region.  We present the 
first simulations of THz quantum well hot-hole lasers that
can produce inversion and optical gain.

\end{abstract}

\maketitle
\thispagestyle{fancy}


\section{Introduction}
\label{intro}

 Hot hole lasers \cite{OQE23-1991,Hovenier-MPSSW-1997apl} emit in the THz
(far-infrared) with an unusually broad gain spectrum, allowing amplification
and generation of laser pulses on a picosecond time scale.  The THz band has
important potential applications in (e.g.) medical imaging and office
communications.  Bulk hot-hole lasers have been realised in p-Ge, producing
gains of $\sim$0.25cm$^{-1}$ around 4THz.  Of the III-V materials, both GaAs
and InSb are suitable for hot hole lasers; although their performance is not as
good as in Ge \cite{Kinsler-W-2000jap-pre}.  Investigation of these is the most
useful for industrial applications because of the existing ability to grow high
quality III-V structures.  

We present a discussion of likely modes of operation of quantum well hot-hole
lasers in GaAs, together with predictions from Monte Carlo simulations for one
design.  The simulations use an infinite-well $k.p$ bandstructure, and include
optical phonons, acoustic phonons, and piezoelectric phonons
\cite{MonteCarlo,Kane-1966ss,Ridley-QPS}.  As yet they do not include ionised
impurity scattering, but this is a small effect and should not affect the
character of the results significantly.



\section{Quantum Wells}
\label{qwells}

\begin{figure}
\resizebox{0.48\textwidth}{!}{%
  \includegraphics{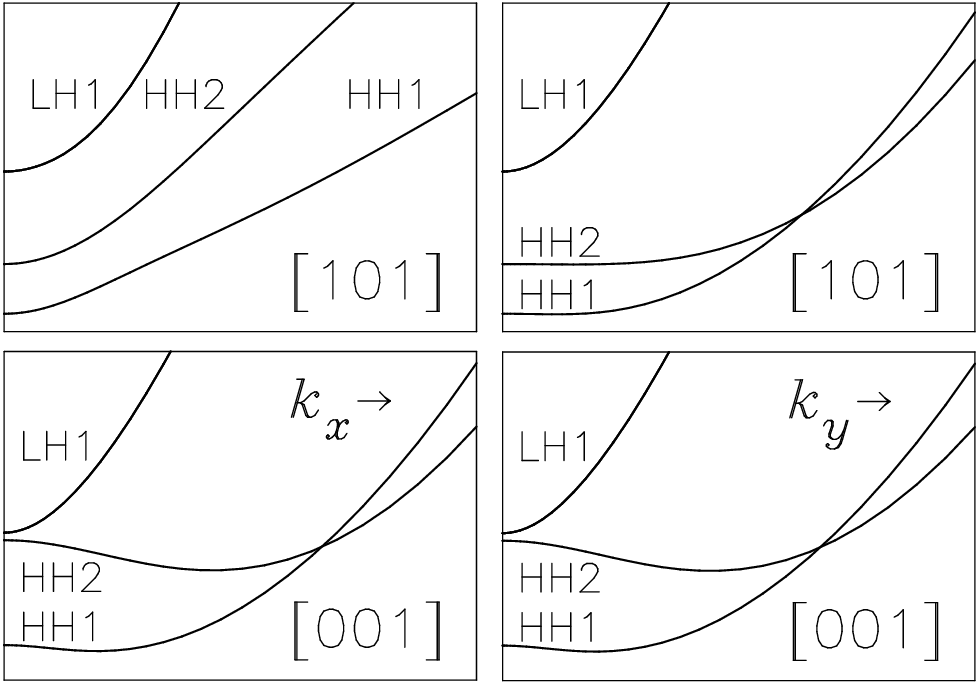}
}
\caption{Band structure of a quantum well hot-hole
laser.  Upper graphs: 100\AA~well with a [101] growth direction; 
lower graphs: a 100\AA~well with a [001] growth direction. 
The $k_x$ variation is shown 
on the left-hand graphs, and $k_y$ on the right. The vertical scale
is 0--100meV, and the horizontal 0--1$\times10^{9}$m$^{-1}$.
}
\label{fig:1} 
\end{figure}

Bulk hot-hole lasers can be described using the heavy and light hole valence
bands: an electric field accelerates the heavy holes in a streaming motion to
high energies $E>E_{LO}$; from where (ideally) they scatter into light hole
cyclotron orbits formed by the magnetic field; and then emit a photon and
return to low energy in the heavy hole band to repeat the cycle.  In a quantum
well each valence band breaks up into a set of subbands: heavy hole subbands
HH1, HH2, HH3, ...; and similarly for light holes LH1, LH2, etc.  The non
parabolicity of the bulk bandstructure leads to a variety of possible quantum
well bandstructures, depending on the orientation of the crystal axes in the
well material.  Figure \ref{fig:1} shows the lower subbands of two simple cases
schematically.  Note that the [101] well is not symmetric in $x$ and $y$; and
that for the [001] well the HH dispersions are no longer even approximately
parabolic, with the HH2 having a noticeable local maximum at the origin.  

The [101] well has two heavy-hole subbands (HH1, HH2) that need to be
considered.  For wells over about $50$\AA~wide, the two lowest energy, and
hence most heavily populated subbands will be HH1 and HH2.
This means that in contrast to the bulk case we do not need a magnetic field to
confine the light holes in cyclotron orbits.  The LH1 is well above the HH2 for
a 100\AA~well, and so has minimal effect on hot-hole laser operation.  Along
one direction ($y$) both HH subbands are very flattened at the base, but with a
LH-like curvature for larger $k_y$; whereas along the other ($x$) HH2 sits
between HH1 and LH1.  This means that the heavy-hole distributions will be
roughly rectangular in outline; and an increasing electric field will shift the
distributions along its direction.  In the $x$ direction the two HH subbands
(nearly) cross at a point about 15meV above the bottom of HH1, allowing for
fast inter-subband scattering.  Holes that get further up HH1 to an energy
equal to $E_{LO}$ above the bottom of HH2 will quickly emit an LO phonon and
return to populate either HH2 or HH1.  Note that HH2 is flat to larger $k_y$
values than HH1, making it relatively easy to get inversion where HH2 is
flat but HH1 is not.

\begin{figure}
\centerline{
  \resizebox{0.40\textwidth}{!}{%
    \includegraphics{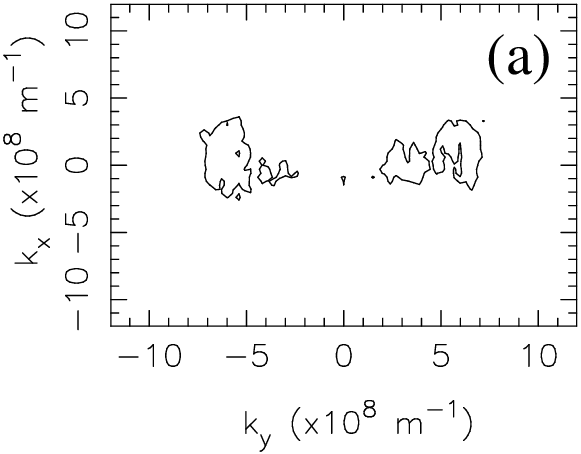}
  }
}
\vspace{3mm}
\centerline{
  \resizebox{0.39\textwidth}{!}{%
    \includegraphics{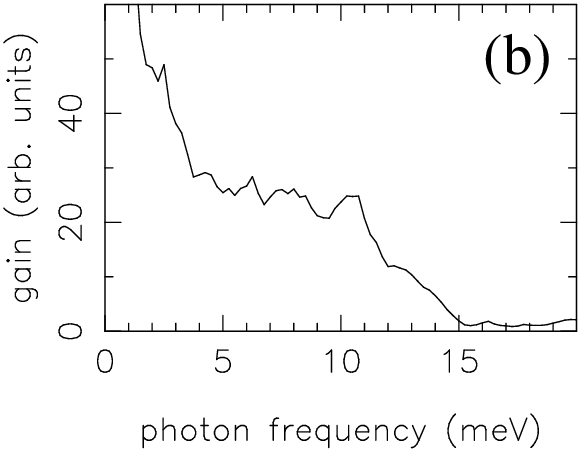}
  }
}
\caption{[101] 100\AA ~quantum well hot-hole laser, with electric field 
$F_x=250$V/cm (a) Contour 
plot showing the region of inversion; (b) spectrum of the gain 
cross-section for $y$ polarised light.  Both graphs are affected by 
statistical noise from the simulations.}
\label{fig:2}
\end{figure}

Fig. \ref{fig:2}(a) shows the difference in the distribution functions between
HH2 and HH1, for an electric field of 250V/cm along the $x$ direction.  We see
regions of inversion for $k_x \approx 0$ and $k_y \approx 5.00 \times
10^{8}$m$^{-1}$.  There is a small displacement in the direction of the field,
and the amount of inversion decreases with increasing field strength.  If the
electric field is applied along the $y$ direction, we see only one inversion
peak, because the other is wiped out as the HH1 distribution is shifted
underneath it; and a field applied along $x=y$ has a similar, but not so
marked, effect.  Figure \ref{fig:2}(b) shows the optical gain due to the
inversion obtained and shown in (a).  We see optical gain occurring over the
range of energies between the small splitting at the HH1-HH2 anti-crossing
($\approx 0$meV) and the separation of the subband minima ($\approx 15$meV).

The [001] well HH2 subband has a distinct local maximum at $k=0$; and also LH1
is rather close to HH2, and should not be neglected.  Our code does not yet
allow for these more complicated dispersions, but we can see that a lasing
cycle might be as follows: A HH1 of moderate positive $k_x$ is accelerated by
an electric field $+F_x$, where it might scatter into HH2 near the
anti-crossing.  Then it will scatter back onto the inverted part of HH2, either
by acoustic phonon scattering or LO phonon emission.  Here it will be
accelerated {\it backwards} past $k=0$ to the point of inflection ($-k_{LHi}$)
where the HH2 effective mass is infinite; and subsequently emit a photon and
drop down to HH1, from where it will eventually scatter to moderate positive
$k_x$ and will repeat the cycle.  Note that HH1 also has a local maxima, but it
is smaller and has less effect; and its point of inflection where holes can
collect is offset from the HH2 one, and so should not interfere with population
inversion too much.  In contrast to the [101] well, we expect the emission
spectrum of this [001] well to be peaked and centered at
$E_{HH2}(k_{LHi})-E_{HH1}(k_{LHi})$, where the HH2's will accumulate.

Quantum well hot hole lasers also benefit from modulation doping. One
significant source of unhelpful scattering is that due to holes scattering off
either other holes or impurities.  We can halve this contribution to the
scattering simply by moving impurities used to add the holes to the device to
outside the active region.  Also, we can vary the well widths to vary the
inter-subband separations to greater or less than $E_{LO}$ to enhance or reduce
optical phonon scattering as required; or even use the well's crystal
orientation to adjust the densities of states (DOS) of the subbands. The [101] well has HH1, HH2 with flat $E(k)$ in the $k_y$ direction,
leading to an enhanced density of states, hence making those regions
into relatively  preferred
destinations for scatterings.  This differs markedly from the DOS effects
in bulk material \cite{Kinsler-W-2000jap-pre}.


\section{Conclusions}
\label{conclusions}

We have shown the potential for quantum well hot-hole lasers by an
investigation of the possible bandstructure in combination with computer
simulation.  Although we have not yet explored the full parameter space of
possible designs, our first attempt (the 100\AA~[101] quantum well) produced
simulations which showed inversion over a range of electric field strengths and
directions, with the optimum being for a 250V/cm field in the in-plane $x$
direction.  Further, the 100\AA~[001]well has a bandstructure which promises a
good inversion.  We aim to continue to test different designs and field
combinations, including the addition of magnetic fields, in order to determine
the most practical designs for experimental investigation.



\vspace{2mm}
\noindent
{\bf Acknowledgements:}\label{acknowledgements}
This work is funded by the European Commission via the 
program for Training and Mobility of Researchers.

\end{document}